\documentclass[sigconf]{acmart}
\AtBeginDocument{%
  \providecommand\BibTeX{{%
    \normalfont B\kern-0.5em{\scshape i\kern-0.25em b}\kern-0.8em\TeX}}}

\setcopyright{acmcopyright}
\copyrightyear{2018}
\acmYear{2018}
\acmDOI{XXXXXXX.XXXXXXX}

\acmConference[Conference acronym 'XX]{Make sure to enter the correct
  conference title from your rights confirmation emai}{June 03--05,
  2018}{Woodstock, NY}
\acmPrice{15.00}
\acmISBN{978-1-4503-XXXX-X/18/06}

\usepackage{multirow}
\usepackage{multicol}
\usepackage{subcaption}
\usepackage{scalefnt}

\usepackage{tikz}
\usetikzlibrary{chains,scopes,positioning,backgrounds,shapes,fit,shadows,calc,arrows.meta}
\tikzset{
    head/.style = {fill = orange!90!blue,
    label = center:\textsf{\Large L}},
    tail/.style = {fill = blue!70!yellow, text = black,
    label = center:\textsf{\large R}}
}
\usepackage{algorithm}
\usepackage{algorithmic}
\usepackage{xcolor}
\usepackage{graphicx}
\usepackage{pgfplots}
\usepgfplotslibrary{groupplots}

\definecolor{mercury}{RGB}{230,230,230}
\definecolor{gallery}{RGB}{240,240,240}
\definecolor{bgc}{RGB}{245,245,245}
\definecolor{tuatara}{RGB}{67, 67, 67}
\definecolor{free_speech_aquamarine}{RGB}{0, 156, 114}
\definecolor{shakespeare}{RGB}{85, 154, 193}
\definecolor{sun_shade}{RGB}{255, 144, 68}
\definecolor{flamingo}{RGB}{237, 88, 85}

\copyrightyear{2023}
\acmYear{2023}
\setcopyright{rightsretained}
\acmConference[KDD '23]{Proceedings of the 29th ACM SIGKDD Conference on
Knowledge Discovery and Data Mining}{August 6--10, 2023}{Long Beach, CA,
USA}
\acmBooktitle{Proceedings of the 29th ACM SIGKDD Conference on Knowledge
Discovery and Data Mining (KDD '23), August 6--10, 2023, Long Beach, CA, USA}
\acmDOI{10.1145/3580305.3599919}
\acmISBN{979-8-4007-0103-0/23/08}

\begin{document}

\title{Tree based Progressive Regression Model for Watch-Time Prediction in Short-video Recommendation}

\author{Xiao Lin}
\authornote{Both authors contributed equally to this research.}
\email{jackielinxiao@gmail.com}
\affiliation{%
  \institution{Kuaishou Technology}
  \city{Beijing}
  \country{China}
}

\author{Xiaokai Chen}
\authornotemark[1]
\email{chenxiaokai@kuaishou.com}
\affiliation{%
  \institution{Kuaishou Technology}
  \city{Beijing}
  \country{China}
}

\author{Linfeng Song}
\email{songlinfeng@kuaishou.com}
\affiliation{%
  \institution{Kuaishou Technology}
  \city{Beijing}
  \country{China}}

\author{Jingwei Liu}
\email{liujingwei@kuaishou.com}
\affiliation{%
  \institution{Kuaishou Technology}
  \city{Beijing}
  \country{China}}

\author{Biao Li}
\email{libiao@kuaishou.com}
\affiliation{%
  \institution{Kuaishou Technology}
  \city{Beijing}
  \country{China}}



\author{Peng Jiang}
\authornote{Corresponding author.}
\email{jiangpeng@kuaishou.com}
\affiliation{%
\institution{Kuaishou Technology}
  \city{Beijing}
  \country{China}
  }


\renewcommand{\shortauthors}{Trovato and Tobin, et al.}

\begin{abstract}
  An accurate prediction of watch time has been of vital importance to enhance user engagement in video recommender systems. To achieve this, there are four properties that a watch time prediction framework should satisfy: first, despite its continuous value, watch time is also an ordinal variable and the relative ordering between its values reflects the differences in user preferences. Therefore the ordinal relations should be reflected in watch time predictions. Second, the conditional dependence between the video-watching behaviors should be captured in the model. For instance, one has to watch half of the video before he/she finishes watching the whole video. Third, modeling watch time with a point estimation ignores the fact that models might give results with high uncertainty and this could cause bad cases in recommender systems. Therefore the framework should be aware of prediction uncertainty. Forth, the real-life recommender systems suffer from severe bias amplifications thus an estimation without bias amplification is expected.
  
  How to design a framework that solves these four issues simultaneously remain unexplored. Therefore we propose TPM (Tree-based Progressive regression Model) for watch time prediction. Specifically, the ordinal ranks of watch time are introduced into TPM and the problem is decomposed into a series of conditional dependent classification tasks which are organized into a tree structure. The expectation of watch time can be generated by traversing the tree and the variance of watch time predictions is explicitly introduced into the objective function as a measurement for uncertainty. Moreover, we illustrate that backdoor adjustment can be seamlessly incorporated into TPM, which alleviates bias amplifications.
  
  Extensive offline evaluations have been conducted in public datasets and TPM have been deployed in a real-world video app Kuaishou with over 300 million DAUs. The results indicate that TPM outperforms state-of-the-art approaches and indeed improves video consumption significantly.
\end{abstract}





\maketitle

\section{Introduction} 
Recent years have witnessed the growing popularity of online video services (e.g. YouTube and Hulu) and video-sharing platforms (TikTok and KuaiShou). And the amount of time that users spend on watching recommended videos (which is referred to as watch time) becomes a key metric reflecting user engagement. Users who get recommendations with higher watch time tend to stay longer in the platform, which brings a growth of DAU (Daily Active User).

Despite its importance, watch time prediction has not been widely studied in previous researches \cite{covington2016deep,zhan2022deconfounding}. We argue that there are several special important aspects in watch time modeling:

First, watch time prediction is essentially a regression problem, but the ordinal relation between watch time predictions is also important in recommendation. On one hand, watch time is a continuous random variable, the recommender system needs to get an accurate prediction of watch time for the usage in downstream phases; on the other hand, watch time is a metric for video comparison and the ordinal relation of predictions is also important. For example, given two predictions of watch time for a video: $T_1 = 3.5s$, $T_2 = 4.5s$ and the ground truth is $T=4s$. The two predictions share same regression error of $0.5s$ in terms of MAE. However these two predictions lead to very different consequences in recommender systems. As the system usually tends to recommend videos with higher predicted watch time, the video is much more likely to be recommended with a prediction of $4.5s$ compared to the case with $3.5s$.
Therefore, estimating watch time with direct regression fails to model the ordinal relations between watch time. Ranking losses focus on the ordinal relations but may lead to predictions that deviate far from the ground truth. Therefore a good formulation of watch time prediction should satisfy both requirements simultaneously.

Second, there exists strong conditional dependence in the video-watching behaviors. For example, one has to watch half of the video before he/she finishes watching the whole video. This is similar to the case of click and post-click behaviors (e.g. purchase) in E-commerce platforms \cite{ma2018entire,wen2020entire}. This conditional dependence needs to be considered in watch time prediction.

Third, to enable a robust prediction of watch time, we expect the prediction model to be uncertainty-aware about its predictions. For most regression models, the objective is to get an accurate point estimation by minimizing a $L_1$ or $L_2$ loss. Thus the models might produce predictions with high uncertainty. For real-life recommender systems, this could lead to bad cases where sub-optimal videos are assigned with high rankings by the model, but cause unsatisfactory user experiences.
However, how to model the uncertainty in watch prediction is still under-investigated. 

\begin{table}
  \caption{Comparison between WLR, D2Q and TPM}
  \label{tab:comp_approaches}
  \begin{tabular}{ccccc}
    \toprule
    Approaches & WLR & D2Q & TPM \\
    \midrule
    Ordinal Relation & $\times$ & $\times$ & $\checkmark$ \\
    Conditional Dependence & $\times$& $\times$ & $\checkmark$ \\
    Model Uncertainty & $\times$ & $\times$ & $\checkmark$ \\
    Debiasing & $\times$ & $\checkmark$ & $\checkmark$\\
  \bottomrule
\end{tabular}
\end{table}

Forth, most real-life recommender systems suffer from bias amplifications (e.g. sample selection bias, popularity bias). As the training data for models is usually collected from the logs in the platform, this causes severe bias amplification. According to previous studies \cite{zhan2022deconfounding}, the recommendation of video recommender systems can be biased towards videos with longer durations, which verifies the existence of bias amplification.

Given the four issues, we have reviewed existing studies on watch time prediction.
Although both methods tackle some important limitations and achieve superior performances in this task, none of them have fully considered all the four issues. The analyses on two state-of-the-art methods on watch time prediction (WLR \cite{covington2016deep} and D2Q \cite{zhan2022deconfounding}) can be found in Table \ref{tab:comp_approaches}. 

In WLR, training samples are either positive (the video impression was clicked) or negative (the impression was un-clicked). And the watch-time prediction is treated as a binary classification problem, where the positive samples are weighted with watch time in the cross-entropy loss. Therefore the odds learned by the classifier equals to the expected watch-time approximately. Despite its simplicity and effectiveness, WLR still has some limitations that prevent its direct application in full-sceen video recommender systems \cite{zhan2022deconfounding}, where all video impressions are watched. Therefore, WLR has to be trained with artificially designated positive and negative samples and weights, which may cause a poor approximation of watch-time. Meanwhile the bias amplification effect may get even more severe in WLR as more weights are assigned to the videos of longer duration.
D2Q \cite{zhan2022deconfounding} alleviates the duration bias by splitting videos into different groups according to their durations and models watch time with traditional regression models in each group. Therefore the ordinal relationships and conditional dependence in watch time values are neglected; Moreover, both WLR and D2Q treats watch time prediction as a point estimation problem, thus the uncertainty of predictions is ignored.

\tikzset{global scale/.style={
    scale=#1,
    every node/.append style={scale=#1}
  }
}

\begin{figure} 
\centering
\scalefont{1.5}
\begin{tikzpicture}
[semithick,->,grow=down, global scale = 0.6,level distance = 1.5cm,level 1/.style = {sibling distance=7.0cm},
level 2/.style = {sibling distance=4.0cm},
level 3/.style = {sibling distance=3.0cm},
sibling distance=6.0cm]
\node [draw,rectangle, line width = 1pt ] (root) {$n_0:\newline [0.0,1.0]$}
child {node [draw,rectangle, line width = 1pt] (A) {$n_1:[0.0,0.6]$}
    child {node [draw,rectangle, line width = 1pt] (C) {$n_3:[0.0,0.2]$}}
    child {node [draw,rectangle, line width = 1pt] (D) {$n_4:[0.2,0.6]$}
    child {node [draw,rectangle, line width = 1pt] (G) {$n_7:[0.2,0.4]$}}
    child {node [draw,rectangle, line width = 1pt] (H) {$n_8:[0.4,0.6]$}}
    }
}
child {node [draw,rectangle, line width = 1pt] (B) {$n_2:[0.6,1.0]$}
    child {node [draw,rectangle, line width = 1pt] (E) {$n_5:[0.6,0.8]$}}
    child {node [draw,rectangle, line width = 1pt] (F) {$n_6:[0.8,1.0]$}}
};

\end{tikzpicture}

\caption{An example of a tree in TPM. Each node is assigned with a classifier deciding which interval the prediction of watch time belongs to. And each edge represents a result of the decision from parent node, which serves as a condition for current node. For instance, the node $(0.2,0.6)$ represents a decision whether the prediction of watch time should go to $(0.2,0.4)$ or $(0.4,0.6)$ given the condition that it belongs to $(0.2,0.6)$}\label{fig:demo_tree}
\end{figure}
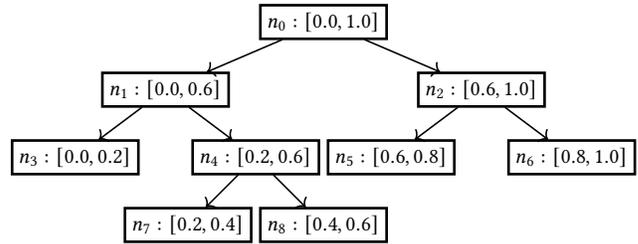
Considering the four aforementioned issues, we propose a new framework TPM for watch time prediction that solves them simultaneously. Specifically, watch time is split into multiple ordinal intervals, and watch time prediction is equivalent to a searching problem deciding which interval the predicted watch time belongs to. The searching process is modeled as a collection of decision making problems organized in a tree structure. Each intermediate node in the tree represents a decision problem which is assigned with a corresponding classifier. Meanwhile each leaf node represents one of the ordinal intervals which is assigned with an expected value of watch time. Each edge represents a possible result of the decision and leads to the next decision. Then the result from upper level becomes the condition for the decisions in the current nodes. Therefore the path from root node to a certain leaf node corresponds to a searching trajectory consisting of a series of decisions. We present a running example in Fig \ref{fig:demo_tree} to illustrate the framework.




We explain how TPM solves the four issues in detail as follows:
\begin{itemize}
    \item First, we introduce ordinal ranks into the approximation of watch time. The regression task is decomposed into multiple binary classification tasks whose labels are associated with the ordinal ranks. This approximation makes use of both the continuity of watch time and the ordinal relations between the ranks. 
    
    \item Second, we introduce the conditional dependence into TPM. Each task of a child node is dependent upon the task from the parent node. In this way, the conditional dependency between the multiple decomposed classification tasks is explicitly modeled. 
    Notice that there are multiple decomposition patterns of the prediction task, we can encode arbitrary conditional dependence into the model. 
    
    \item Third, to enable a robust framework for watch time prediction, we introduce model uncertainty into the objective function. Thanks to the splitting of watch time into ranks, the predicted watch time can be seen as a random variable drawn from multinomial distribution, and we can compute the variance of predicted watch time explicitly, which can be seen as a metric of model uncertainty. Therefore it is introduced into the objective function so that TPM gets an accurate estimation of watch time with high confidence.

    \item Forth, we conduct a causal analysis on the confounding effect of biases and show that conducting backdoor adjustment is equivalent to a specific decomposition with multiple classification tasks. We show that this method applies to different kinds of biases and D2Q \cite{zhan2022deconfounding} can be seen as one of the special cases of TPM.
\end{itemize}

Notice that the structure of TPM is similar to decision trees, we argue that it possesses significant differences with conventional tree models for regression:
First, TPM uses the tree structure to decompose a pure regression problem into a series of classification problems, thus the decomposition is conducted on the label space (e.g. the partition of watch time intervals); while tree models partition the feature space into sub-spaces for prediction. 
Second, the tree-alike decomposition in TPM assigns each node with a corresponding classifier (like neural networks) for decision making; while conventional tree models directly learn a feature partition rule for labeling.

The contributions of this paper are summarized as follows:
\begin{itemize}
    \item We propose a new framework TPM for watch-time prediction, transforming regression problem into a series of multiple classification problems; And both ordinal relation and conditional dependence are introduced into TPM, which are neglected in previous studies;
    \item To enhance the robustness of TPM, we model watch time as a random variable drawn from a multinomial distribution and introduce variance into the objective function for uncertainty modeling in training. 
    \item To the best of our knowledge, TPM is the first approach that considers ordinal relationships, conditional dependence and model uncertainty in watch time prediction.
    \item We show that TPM can be easily adapted for debiased recommendation, which enables easy backdoor adjustment for alleviating bias amplifications;
    \item Extensive experiments are conducted in both offline datasets and a real-world short video recommender system i.e. Kuaishou APP. TPM has achieved an improvement of more than $0.2\%$ in watch time and a significant drop of negative feedbacks.  And TPM has already been deployed online, serving more than $300$ million DAUs.
\end{itemize}

\section{Related Work}
\label{related_work}
\subsection{Watch-time Prediction}
Watch-time prediction is one of the most-concerned problems in industrial recommender systems (especially for short-video and movie recommender systems). 
However, to the best of our knowledge, only few papers can be found\cite{covington2016deep,zhan2022deconfounding} in this area. The first work \cite{covington2016deep} focused on video recommendation in Youtube and proposed the Weighted Logistic Regression (i.e. WLR) method for watch-time prediction. And WLR has become a state-of-the-art method in related applications. However, this method can not be directly applied to full-screen video recommender systems and WLR may suffer from severe bias issues due to its weighting mechanism. D2Q \cite{zhan2022deconfounding} alleviates the duration bias by conducting backdoor adjustments and models watch time with direct watch-time quantile regression. However the ordinal relationships and dependency between quantiles are ignored in this method. Moreover, as both methods model watch time with point estimation, the uncertainty of predictions have not been considered.

\subsection{Ordinal regression}
Ordinal regression is a technique for predicting ordinal labels, i.e. the relative order of labels is important. Its application can be found in age estimation \cite{niu2016ordinal}, monocular depth estimation \cite{fu2018deep}, head-pose estimation \cite{hsu2018quatnet}. Despite the wide applications of ordinal regression, it has not been applied on watch time prediction tasks.

Most ordinal regression algorithms are modified from classical classification algorithms. For instance, SVM has been incorporated with multiple thresholds and applied on visual classification \cite{NIPS2002_51de85dd}; another example is a combination with Online perceptron algorithm \cite{NIPS2001_5531a583} which is used for rating prediction. Moreover, ordering information in class attributes is exploited for transforming ordinal regression into multiple classification problems \cite{frank2001simple}. It is worth noticing that decision tree is used in this work \cite{frank2001simple}. However, the binary classification problems in ordinal regression are not conditional dependent as those in TPM, which is a fundamental difference.

\subsection{Tree based neural networks for recommendation}
Tree based models and neural networks are powerful models in various machine learning applications, especially in recommender systems. Tree based methods like LambdaMART\cite{burges2010ranknet} are fairly competitive in ranking tasks. Meanwhile, neural networks achieve state-of-the-art performances in leveraging sparse and complex features. However, few efforts have been made on combing the advantages of both methods. An early study \cite{li2019combining} attempts to combine decision trees with neural networks for searching. Two models are combined with ensemble learning techniques (e.g. linear combination and stacking) and the combined model achieves superior performances over single models. Moreover, tree models have been used for enhancing the embedding models for explainable recommendation \cite{wang2018tem}.

TDM \cite{zhu2018learning,zhu2019joint} is an example of combining tree-based models and neural networks for recommendation. The idea of TDM is to organize the candidate retrieval process as a searching process along the tree, so that most preferred candidates can be retrieved with arbitrary complex models in logarithmic complexity. TPM differs from TDM in several important aspects: first, TPM is designed for watch time prediction given users and corresponding videos, while TDM aims to retrieve relevant candidates from a huge corpus; second, TPM uses tree structure for problem decomposition while TDM utilizes tree structure for corpus partition; third, TPM traverse the tree to predict expected watch time while TDM uses beam search to search for the target leaf nodes.
\subsection{Debiased recommendation}
Many efforts have been drawn to address the biases in recommendation. Previous studies on this topic can be roughly divided into three categories: 
\begin{itemize}
    \item inverse propensity scoring: it first computes the propensity score of samples based on certain assumptions and then the samples are weighted with inverse propensity scores in the objective function. For example, exposure propensity has been utilized for solving miss-not-at-random problem\cite{saito2020unbiased}; However the performances of this method are sometimes unstable due to the high variance of the estimated propensity. And this can be alleviated by propensity clipping or doubly-robust learning \cite{wang2019doubly} via data imputation. 
    \item causal embedding: this method \cite{bonner2018causal} decomposes related embeddings into unbiased component and biased component. Both components are used at the training stage and biased component is discarded at inference stage to get an unbiased prediction.
    \item causal intervention: the causes of biases are introduced into the method and interventions are conducted to eliminate their affections on recommendation. Randomization and backdoor adjustment are two representative methods for causal intervention. However, it is costly to conduct randomized experiments on real-life recommender systems, and backdoor adjustment \cite{zhan2022deconfounding,wang2021deconfounded} is preferred in practical scenarios.
\end{itemize}
Causal intervention method has been used in watch time prediction for deconfounding duration bias. And we show that backdoor adjustment can be seamlessly incorporated into TPM and this method applies to other confounding factors.

\section{Tree based Progressive regression Model} 
\label{TPM_intro}
We first provide a general formulation for Tree based Progressive regression Model and introduce how watch time prediction is decomposed into several conditional dependent classification problems; Then we present the details of uncertainty modeling for TPM; After that we show that the backdoor adjustment naturally fits in TPM and how biase amplifications are alleviated in detail.
Before we go deep into the details of the formulation, we provide a list of notations in Table \ref{tab:notations}.

\begin{table}
  \caption{Notations}
  \label{tab:notations}
  \begin{tabular}{ccl}
    \toprule
    Notation & Meaning\\
    \midrule
    $T$ & Watch time\\
    $X$ & Features of user and video\\
    $\mathcal{T}$ & Decomposition tree in TPM\\
    $N_{\mathcal{T}}:\{n_i\}$ & set of nodes in $\mathcal{T}$ \\
    $L_{\mathcal{T}}:\{l_k\}$ & set of leaf nodes in $\mathcal{T}$\\
    $\mathcal{M}_i$ & the classifier assigned to $n_i$\\
    $D$ &confounding factor in causal graph\\
    $\phi_{l_k}$ & the path from root to leaf node $l_k$\\
    $n_{\phi_{l_k}(i)}$ & the node at level i along path $\phi_{l_k}$\\
    $d(l_k)$ & the depth of leaf node $l_k$\\
    $\gamma_i,\forall i$ & ordinal ranks of watch time \\
  \bottomrule
\end{tabular}
\end{table}

\subsection{Formulation for TPM}




Given a training instance $(X,T)$ where $X$ represents the features of user and video, and $T$ represents the watch time, the purpose of watch time prediction is to find a model $\mathcal{M}$ so that the prediction $\mathcal{M}(X)$ is close to $T$ under certain metrics.

Instead of treating the problem as direct regression, we first quantize the scale into ordinal ranks: $\{\gamma_0 \leq \gamma_1,\ldots,\gamma_k,\ldots,\leq \gamma_m\}$, and then cast watch time prediction as the estimation of expected ordinal rank. And the estimation from ordinal ranks is similar to a searching process with iterative comparisons.

For instance, if we conduct a linear search along the ranks, the searching process is as follows:
we first decide if $t(x)\leq \gamma_0$ or not: if $t(x)\leq\gamma_0$, then the predicted ordinal rank is $\gamma_0$; otherwise we continue by deciding if $t(x)\leq\gamma_1$, if it is true, then the rank is $\gamma_1$, otherwise the process continues. And the process goes on until an interval is finally found.

And if we conduct a binary search, the searching process then becomes:
we first decide if $t(x)\leq\gamma_{m/2}$: if it is true, then $t(x)$ falls into the set $\{\gamma_0, \gamma_1,\ldots,\gamma_{m/2}\}$; otherwise it belongs to a rank from $\{\gamma_{m/2+1}, \ldots,\gamma_{m}\}$. The process continues and the searching space is narrowed down to a certain rank.

\tikzset{global scale/.style={
    scale=#1,
    every node/.append style={scale=#1}
  }
}

\begin{figure} 
\captionsetup[subfigure]{font=footnotesize}
\subcaptionbox{An unbalanced binary tree for linear search}[0.45\textwidth]
{%
\scalefont{1.8}
\begin{tikzpicture}[semithick,->,grow=down, global scale = 0.5,level distance = 2.2cm,sibling distance=4.0cm]
\node [draw,rectangle, line width = 1pt] (root) {$n_0:[0.0,1.0]$}
child {node [draw,rectangle, line width = 1pt] (A) {$n_1:[0.0,0.25]$}}
child {
node [draw,rectangle, line width = 1pt] (B) {$n_2:[0.25,1.0]$}
child {node[draw, rectangle, line width = 1pt] (C) {$n_3:[0.25,0.5]$}}
child {node[draw,rectangle, line width = 1pt] (D) {$n_4:[0.5,1.0]$}
child {node[draw, rectangle, line width = 1pt] (E){$n_5:[0.5,0.75]$}}
child {node[draw,rectangle, line width = 1pt] (F){$n_6:[0.75,1.0]$}}
}
};

\begin{scope}[nodes={draw=none}]
\path (root) -- (B) node [near start, right] {$\mathcal{M}_0:p(T\in n_2|n_0)$};
\path (B) -- (D) node [near start, right] {$\mathcal{M}_2:p(T\in n_4|T\in n_2)$};
\path (D) -- (F) node [near start, right] {$\mathcal{M}_4:p(T\in n_6|T\in n_4)$};
\end{scope}
\end{tikzpicture}}

\tikzset{global scale/.style={
    scale=#1,
    every node/.append style={scale=#1}
  }
}
\subcaptionbox{A balanced binary tree for binary search}[0.45\textwidth]{
\scalefont{1.8}
\begin{tikzpicture}[semithick,->,grow=down,global scale = 0.5,level distance = 3.0cm,
level 1/.style = {sibling distance=6.5cm},
level 2/.style = {sibling distance=3.0cm}
]
\node [draw,rectangle, line width = 1pt,align=center] (root) {$n_0:$\\$[0.0,1.0]$}
child {node [draw,rectangle, line width = 1pt,align=center] (A) {$n_1:$\\$[0.0,0.5]$}
    child {node [draw,rectangle, line width = 1pt,align=center] (C) {$n_3:$\\$[0.0,0.25]$}}
    child {node [draw,rectangle, line width = 1pt,align=center] (D) {$n_4:$\\$[0.25,0.5]$}}
}
child {node [draw,rectangle, line width = 1pt,align=center] (B) {$n_2:$\\$[0.5,1.0]$}
    child {node [draw,rectangle, line width = 1pt,align=center] (E) {$n_5:$\\$[0.5,0.75]$}}
    child {node [draw,rectangle, line width = 1pt,align=center] (F) {$n_6:$\\$[0.75,1.0]$}}
};

\begin{scope}[nodes={draw=none}]
\path (root) -- (B) node [near start, right] {$\mathcal{M}_0:p(T\in n_2|T\in n_0)$};
\path (A) -- (D) node [near start, right] {$\mathcal{M}_1:p(T\in n_4|T\in n_1)$};
\end{scope}

\end{tikzpicture}}
\caption{Two examples of decomposition trees in TPM}\label{fig:trees_in_tpm}
\end{figure}
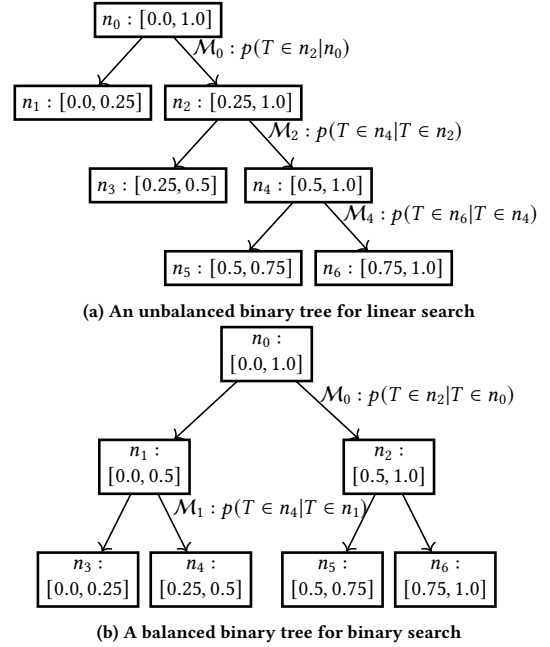

Notice that each searching process consists of a sequence of decisions, we propose to fit the watch time prediction model into a searching process from root to a leaf node along the tree. (see Fig. \ref{fig:trees_in_tpm}): for linear search case, the tree is an unbalanced binary tree; for binary search case, it is a balanced binary tree.

Therefore we propose a Tree based Progressive Model (TPM) for watch time prediction. The model consists of a tree for problem decomposition $\mathcal{T}$ and corresponding classification models $\{\mathcal{M}_i, i\in \{0,1,\ldots, |N_{\mathcal{T}}|-|L_{\mathcal{T}}|\}\}$ where $N_{\mathcal{T}}$ is the set of nodes in $\mathcal{T}$ and $L_{\mathcal{T}}$ is the set of leaf nodes in $\mathcal{T}$(see Fig \ref{fig:trees_in_tpm} as an overview).

The tree in TPM consists of a set of nodes: $\mathcal{T(X)}=\{N_{\mathcal{T}}\bigcup L_{\mathcal{T}}\}$. Each non-root node represents an interval consisting of consecutive ordinal ranks, i.e. $n_{i}: [\gamma_{s_i}, \gamma_{e_i}],e_i-s_i>1$. Without loss of generality, root node is assumed to be the full space: $T\in [\gamma_0,\gamma_m]$. Each leaf node is assigned with a specific ordinal rank i.e. $l_i: [\gamma_{i},\gamma_{i+1}]$. And the subspace of a parent node is the union of sub-spaces of its children. And a path from root to a leaf $l_k$ is denoted as an ordered node set $\phi_{l_k}=\{\hat{n}_{\phi_{l_k}}(0),\ldots, \hat{n}_{\phi_{l_k}}(d(l_k))\}$, where $\hat{n}_{\phi_{l_k}}(i)$ is the node at level $i$ along the path $\phi_{l_k}$ and $d(l_k)$ is the depth of leaf node $l_k$. The following equation always holds:
\begin{displaymath}
    \hat{n}_{\phi_{l_k}}(i) \subseteq \hat{n}_{\phi_{l_k}}(j), \forall i\geqq j
\end{displaymath}

In TPM, each non-leaf node is assigned with a classifier, and its outputs indicate the conditional probabilities that watch time belongs to the corresponding ordinal ranks of the child nodes given the probability from its parent's output. Therefore given an instance $X$ and a tree $\mathcal{T}$, its predicted watch time $T$ follows a multinomial distribution as follows:
\begin{equation}\label{eqn:loglikelihood}
\begin{aligned}
    p(T\in l_k |X,\mathcal{T}) &= p(T\in \hat{n}_{\phi_{l_k}}(i), \forall i\leq d(l_k) |X, \mathcal{T})\\
    &=p(T\in \hat{n}_{\phi_{l_k}}(d(l_k)) |X, \mathcal{T}, T\in \hat{n}_{\phi_{l_k}}(i),\forall i<d(l_k))\\ &\cdot p(T\in \hat{n}_{\phi_{l_k}}(i), \forall i< d(l_k) |X, \mathcal{T})\\
    &=\prod_{1\leq i\leq d(l_k)}p(T\in \hat{n}_{\phi_{l_k}}(i) |X, \mathcal{T}, T\in \hat{n}_{\phi_{l_k}}(i-1))
\end{aligned}
\end{equation}
where $p(T\in \hat{n}_{\phi_{l_k}}(i) |X, \mathcal{T}, T\in \hat{n}_{\phi_{l_k}}(i-1))$ is parameterized by the classifier $\mathcal{M}_{\hat{n}_{\phi_{l_k}}(i-1)}$ assigned to node $\hat{n}_{\phi_{l_k}}(i-1)$.
Based on the derivations, the expectation of watch time given a tree $\mathcal{T}$ can be computed as follows:
\begin{equation}\label{eqn:expected_watchtime_tree}
    E(T|X,\mathcal{T}) = \sum_{l_k\in L_{\mathcal{T}}} E(T|T\in l_k,X,\mathcal{T}) p(T\in l_k |X,\mathcal{T})
\end{equation}

Notice that the expectation involves the term $E(T|T\in l_k,X,\mathcal{T})$, this can be estimated by any predictive model. In this paper, we employ a simple method for estimation:
\begin{displaymath}
    E(T|T\in l_k,X,\mathcal{T}) = (\gamma_{l_k}+\gamma_{l_{k+1}})/2
\end{displaymath}

By building multiple trees for watch time estimation, the expectation of watch time can be computed by incorporating the distribution of trees as a prior:

\begin{equation}\label{eqn:expected_watchtime}
    E(T|X) = \sum_{\mathcal{T}} E(T|X,\mathcal{T})p(\mathcal{T}|X)=\sum_{\mathcal{T}} E(T|X,\mathcal{T})p(\mathcal{T})
\end{equation}

Despite that TPM allows a bagging scheme for prediction as Eqn. \ref{eqn:expected_watchtime}, we restrict the number of trees to be one for simplicity.

\subsection{Tree Construction}
Notice that there is no limit on the type of trees in TPM, the structure of the trees can be designed according to the tasks and dataset. In TPM, each tree corresponds to a decomposition of the ordinal ranks and each non-leaf node inside the tree corresponds to a classifier. As revealed in previous studies \cite{kaur2019systematic,johnson2019survey}, label imbalance adds difficulty to the predictive modeling, we try to construct the tree with balanced label distribution for each node. 

Therefore, we compute the quantiles of watch time and set them as the ordinal ranks for discretion. Then we split the ordinal ranks into halves iteratively and set them to the leaf nodes of a complete binary tree. The tree is constructed by merging two child nodes to a parent node repeatedly. Therefore the classifier for each node is a binary classifier and the label distribution is balanced for each classifier. For example, when the scale of watch time is split into 4 intervals uniformly, the constructed tree corresponds to the balanced binary tree in Fig \ref{fig:trees_in_tpm}.

\subsection{Uncertainty Modeling}
Previous methods \cite{covington2016deep,zhan2022deconfounding} focus on modeling watch time with point estimation, however it is unknown how much confidence should be placed on the predictions. And we show that TPM does not only model the error of expected watch time but also attempts to minimize the uncertainty of its predictions.

\begin{figure}
\captionsetup[subfigure]{font=footnotesize}
\centering
\subcaptionbox{Prediction from $\mathcal{M}_a$}[.25\textwidth]{%
\begin{tikzpicture}[global scale = 0.55]
\begin{axis}[
legend pos=south east,
x=20pt,
xlabel = {$\gamma$},
ylabel = {$P(T=\gamma|X,\mathcal{T})$},
xmin = 0.5, xmax=9,
ymin=-0.05, ymax=0.3,
xtick={1,2,3,4,5,6,7,8},
ytick={0.1,0.2,0.3,0.4},
xticklabels={1,2,3,4,5,6,7,8}, 
yticklabels={0.1,0.2,0.3,0.4},
line width=1pt,
]
\addplot+[ybar,fill=flamingo,draw=none,mark=none]
	coordinates {(1,0.15) (2.0,0.1)
		 (3,0.05) (4.0,0.2) (5,0.2) (6.0,0.05) (7,0.1) (8.0,0.15)};
\addplot[draw=black,line width = 0.5pt]
	coordinates {(1,0.15) (2.0,0.1)
		 (3,0.05) (4.0,0.2) (5,0.2) (6.0,0.05) (7,0.1) (8.0,0.15)};
\end{axis}
\end{tikzpicture}}%
\subcaptionbox{Prediction from $\mathcal{M}_b$}[.25\textwidth]{
\begin{tikzpicture}[global scale = 0.55]
\begin{axis}[
legend pos=south east,
x=20pt,
xlabel = {$\gamma$},
ylabel = {$P(T=\gamma|X,\mathcal{T})$},
xmin = 0.5, xmax=9,
ymin=-0.05, ymax=0.3,
xtick={1,2,3,4,5,6,7,8},
ytick={0.1,0.2,0.3,0.4},
xticklabels={1,2,3,4,5,6,7,8}, 
yticklabels={0.1,0.2,0.3,0.4},
line width=1pt,
]
\addplot+[ybar,fill=sun_shade,draw=none,mark=none]
	coordinates {(1,0.05) (2,0.1)
		 (3,0.1) (4,0.25) (5,0.25) (6,0.1) (7,0.1) (8,0.05)};
\addplot[draw=black,line width = 0.5pt]
	coordinates {(1,0.05) (2,0.1)
		 (3,0.1) (4,0.25) (5,0.25) (6,0.1) (7,0.1) (8,0.05)};
\end{axis}
\end{tikzpicture}
}
\caption{An example of watch time predictions as a distribution}\label{fig:variance_demo}
\end{figure}
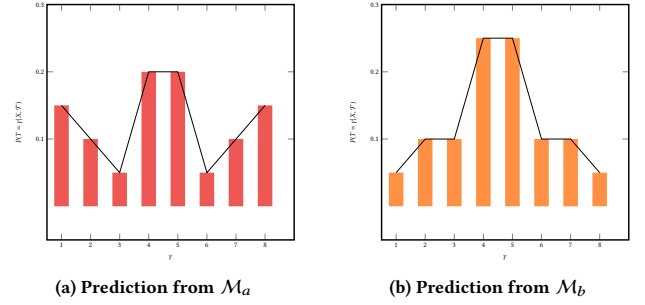

Notice that given a tree for problem decomposition, TPM predicts the probabilities that watch time belongs to the ordinal ranks respectively. Therefore watch time becomes a random variable following a multinomial distribution of $p(T\in l_k |X,\mathcal{T}), \forall l_k\in L_{\mathcal{T}}$.
This property of predicting watch time with a distribution is very helpful, as it enables an approximate estimation of watch time variance:
\begin{equation}\label{eqn:var_watchtime_tree}
    Var(T|X,\mathcal{T}) = E(T^2|X,\mathcal{T}) - E(T|X,\mathcal{T})^2
\end{equation}
Notice that $p(T\in l_k|X,\mathcal{T})$ can be computed with Eqn \ref{eqn:loglikelihood}, the variances can be computed easily under the assumption $T\sim p(T\in l_k |X,\mathcal{T}), \forall l_k\in L_{\mathcal{T}}$.

A simple example is depicted in Fig \ref{fig:variance_demo} to illustrate the idea: assuming that the scale of watch time is split into eight ordinal ranks, the predictions of two models $\mathcal{M}_a$ and $\mathcal{M}_b$ have same expectation of watch time: $E(T)=4.5$. However, it is easy to verify that these two predictions have distinct variances: $Var_{\mathcal{M}_a}(T)>Var_{\mathcal{M}_b}(T)$. This indicates that $\mathcal{M}_a$ is more uncertain about its prediction. And it is expected that a model is able to get correct estimation of watch time with high certainty. Therefore we explicitly add variance of predicted watch time into the objective function of TPM.

\subsection{Training with TPM}
Now we present the training process of TPM. 
Given a training sample $(X,T)$ and the tree $\mathcal{T}$ in TPM, we first identify the ordinal rank of the sample and the corresponding leaf node $l_k(T)\in L_{\mathcal{T}}$. Then the path from root to the leaf node is identified and the sample is associated to the classifiers along the path. 

Each classifier takes $X$ as input, and the label is identified with the child node along the path. In this paper, $\mathcal{T}$ is a balanced binary tree, and each non-leaf node is assigned with a binary classification task. For each classifier, samples belonging to the right-hand child node is seen as a positive sample.

Consider an example in Fig \ref{fig:trees_in_tpm}, given a sample $(X,T)$ and $T = 0.8$, this sample is associated to classifiers $\mathcal{M}_0$ and $\mathcal{M}_2$, and for both classifiers, it is a positive sample.


\begin{algorithm}[t]
\caption{\textbf{Tree based Progressive-regression Model:}}
\label{alg:TPM_tree}
\begin{algorithmic}[1]
\STATE{Input:} Training data: $(X_i, T_i),\forall i$, A decomposition tree: $\mathcal{T}$;
\STATE{Output:} The classifiers of nodes $\mathcal{M}_j,\forall j\in N_{\mathcal{T}}\setminus L_{\mathcal{T}}$;
\FOR{each batch}
\STATE Assign each training sample into the leaf nodes of $\mathcal{T}$ by fitting $T_i$ to the ordinal ranks of $l_k,\forall k$;
\STATE Assign $(X_i,T_i)$ to the classifiers along corresponding path $l_k$;
\STATE Compute the loglikelihood of $(X_i,T_i)$ belonging to path $l_k$ as Eqn. \ref{eqn:loglikelihood};
\STATE Compute $E(T|X,\mathcal{T})$ and $Var(T|X,\mathcal{T})$ as Eqn. \ref{eqn:expected_watchtime_tree} and Eqn. \ref{eqn:var_watchtime_tree};
\STATE Compute the final objective function $\mathcal{L}(T,X,\mathcal{T})$ as Eqn. \ref{eqn:objective_function};
\STATE Update $\mathcal{M}_j,\forall j$ by minimizing $\mathcal{L}(T,X,\mathcal{T})$;
\ENDFOR
\end{algorithmic}
\end{algorithm}

The objective function of TPM consists of three components:
\begin{itemize}
    \item Classification error of classifiers along the path: TPM attempts to maximize the likelihood w.r.t. $p(\hat{T}\in l_k(T) |X,\mathcal{T})$, where $\hat{T}$ is the predicted watch. 
    \item Predicition Variance: $Var(\hat{T}|X,\mathcal{T})$. For easier optimization, we use standard deviation in the loss function: $Var(\hat{T}|X,\mathcal{T})^{0.5}$
    \item Regression error: a loss function evaluating the difference between the final prediction of watch time and groundtruth: $|T-E(\hat{T})|$.
\end{itemize}

The final objective function is a weighted sum of the three components:

\begin{equation}\label{eqn:objective_function}
max.\mathcal{L} = \alpha_1 log(p(\hat{T}\in l_k(T) |X,\mathcal{T}))-\alpha_2 Var(\hat{T}|X,\mathcal{T})^{0.5} - \alpha_3 \|E(\hat{T})-T\|_2
\end{equation}

The training process is illustrated in Alg. \ref{alg:TPM_tree}.






\subsection{Combined with Backdoor Adjustment}
\begin{figure}
  \includegraphics[width=0.25\textwidth]{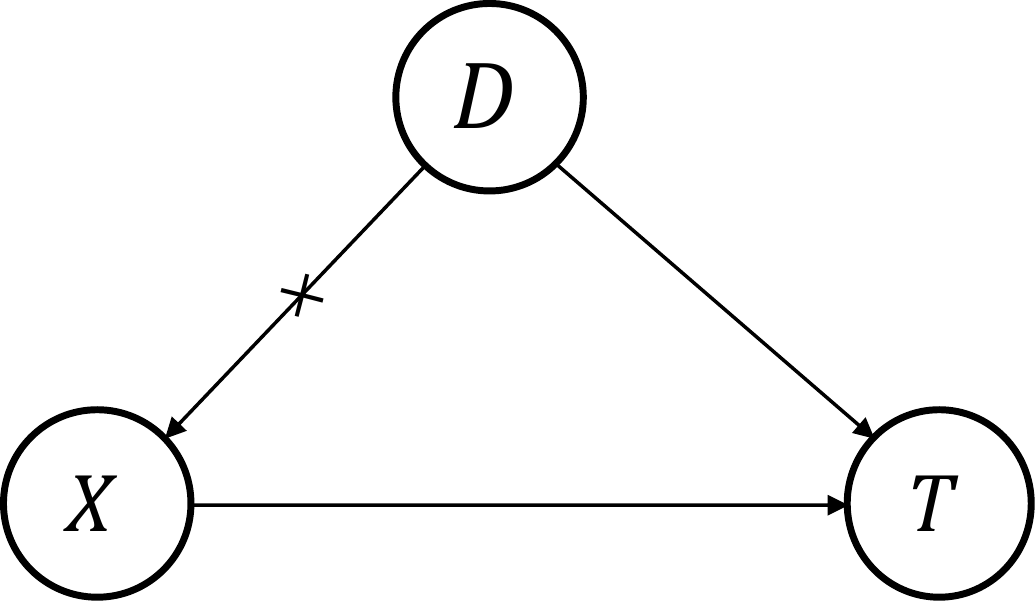}
 \caption{Causal Graph illustrating the confounding effect in watch time prediction. $D$, $X$, $T$ represent the confounding factor, the input features and watch time respectively.}.
  \label{fig:causal_graph_tpm}
\end{figure}

Now we present how backdoor adjustment seamlessly adapts to TPM for debiasing recommendation. First, we present the causal graph in Fig \ref{fig:causal_graph_tpm} to illustrate the bias effects to watch time prediction.
Denote the confounding factor as $D$, the feature representations as $X$ and the watch time as $T$, the effect between variables are reflected in the edges:
\begin{itemize}
    \item $D\rightarrow T$: Confounding factors affect watch time directly. This should be captured by models for an accurate estimation \cite{wang2021deconfounded,zhan2022deconfounding}.
    \item $D\rightarrow X$: Confounding factors affect feature representations implicitly. This should be eliminated so that bias amplifications can be avoided.
    \item $X\rightarrow T$: Feature representations directly affect watch time, including the effects of user preferences and video contents, etc.
\end{itemize}

\tikzset{global scale/.style={
    scale=#1,
    every node/.append style={scale=#1}
  }
}

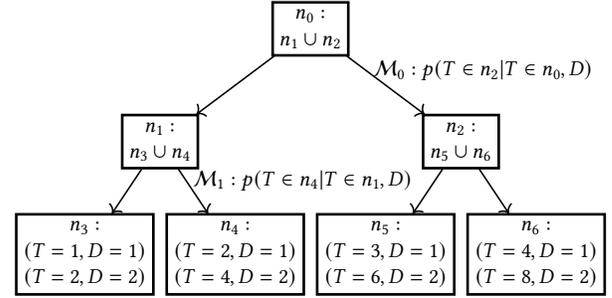
\begin{figure} 
\captionsetup[subfigure]{font=footnotesize}
{%
\scalefont{1.8}
\begin{tikzpicture}[semithick,->,grow=down,global scale = 0.5,level distance = 3.0cm,
level 1/.style = {sibling distance=8.0cm},
level 2/.style = {sibling distance=4.0cm}
]
\node [draw,rectangle, line width = 1pt,align=center] (root) {$n_0:$\\$n_1\cup n_2$}
child {node [draw,rectangle, line width = 1pt,align=center] (A) {$n_1:$\\$n_3\cup n_4$}
    child {node [draw,rectangle, line width = 1pt,align=center] (C) {$n_3:$\\$(T=1,D=1)$\\$(T=2,D=2)$}}
    child {node [draw,rectangle, line width = 1pt,align=center] (D) {$n_4:$\\$(T=2,D=1)$\\$(T=4,D=2)$}}
}
child {node [draw,rectangle, line width = 1pt,align=center] (B) {$n_2:$\\$n_5\cup n_6$}
    child {node [draw,rectangle, line width = 1pt,align=center] (E) {$n_5:$\\$(T=3,D=1)$\\$(T=6,D=2)$}}
    child {node [draw,rectangle, line width = 1pt,align=center] (F) {$n_6:$\\$(T=4,D=1)$\\$(T=8,D=2)$}}
};

\begin{scope}[nodes={draw=none}]
\path (root) -- (B) node [near start, right] {$\mathcal{M}_0:p(T\in n_2|T\in n_0,D)$};
\path (A) -- (D) node [near start, right] {$\mathcal{M}_1:p(T\in n_4|T\in n_1,D)$};
\end{scope}

\end{tikzpicture}}

\caption{An examples of decomposition tree in TPM when backdoor adjustment is conducted. Each node is asscoiated with both watch time ($T$) and the condounding factor $D$.}\label{fig:trees_in_tpm_deconfounding}
\end{figure}

Without loss of generality, we assume $D$ follows a multinomial distribution: $D\sim P(D=d),\forall d$. The deconfounded estimation of watch time can be derived by blocking the edge $D\rightarrow X$ \cite{wang2021deconfounded,zhan2022deconfounding}:
\begin{align}
    E(T|do(X))
    &= \sum_{d}P(D=d|do(X))E(T|do(X),D=d) \\
    &= \sum_{d}P(D=d|X)E(T|X,D=d)\\
    &= \sum_{d}P(D=d)E(T|X,D=d)
\end{align}

Notice that
\begin{displaymath}
    E(T|X) = \sum_{\mathcal{T}} E(T|X,\mathcal{T})p(\mathcal{T}|X)
\end{displaymath}
\begin{displaymath}
    E(T|X,D=d)= \sum_{\mathcal{T}} E(T|X,\mathcal{T},D=d)p(\mathcal{T}|X,D=d)
\end{displaymath}

we have:

\begin{align}
    &E(T|do(X))\\
    &= \sum_{d}P(D=d)E(T|X,D=d)\\
    &= \sum_{d}P(D=d) \sum_{\mathcal{T}} E(T|X,\mathcal{T},D=d)p(\mathcal{T}|X,D=d)\\
    &= \sum_{\mathcal{T}}\sum_{d}P(D=d) E(T|X,\mathcal{T},D=d)p(\mathcal{T}|X,D=d)
\end{align}

Specifically, this indicates that we can conduct backdoor adjustment by constructing trees according to the distribution of confounding factors and train the classifiers by splitting samples to the corresponding trees.
This can be achieved by splitting the scale of confounding factors into groups and construct the tree accordingly. Meanwhile, the training data should be split according to the groups and the classifiers in each group is trained with the split data respectively (See Fig. \ref{fig:trees_in_tpm_deconfounding} for example).

Specifically, we can inject $D$ into TPM as follows:
\begin{equation}\label{eqn:debiased_loglikelihood}
\begin{aligned}
    \mathcal{L}_1 &=p(T\in l_k |X,D,\mathcal{T}) \\
    &=\prod_{1\leq i\leq d(l_k)}p(T\in \hat{n}_{\phi_{l_k}}(i) |X,D, \mathcal{T}, T\in \hat{n}_{\phi_{l_k}}(i-1))
\end{aligned}
\end{equation}
\begin{equation}\label{eqn:debiased_expected_watchtime_tree}
    E(T|X,D,\mathcal{T}) = \sum_{l_k\in L_{\mathcal{T}}} E(T|T\in l_k,X,D,\mathcal{T}) p(T\in l_k |X,D,\mathcal{T})
\end{equation}
\begin{equation}\label{eqn:debiased_var_watchtime_tree}
    \mathcal{L}_2 = Var(T|X,D,\mathcal{T}) = E(T^2|X,D,\mathcal{T}) - E(T|X,D,\mathcal{T})^2
\end{equation}
\begin{equation}\label{eqn:debiased_objective_function}
\mathcal{L}(T,X,D,\mathcal{T}) = \alpha_1 \mathcal{L}_1-\alpha_2 \mathcal{L}_2 - \alpha_3 \|E(\hat{T})-T\|_2
\end{equation}

\begin{algorithm}[t]
\caption{\textbf{Training TPM with Backdoor Adjustment}}
\label{alg:TPM_tree_debiased}
\begin{algorithmic}[1]
\STATE{Input:} Training data: $(X_i, T_i),\forall i$, A confounding factor $D$, A decomposition tree $\mathcal{T}$;
\STATE{Output:} The classifiers of nodes $\mathcal{M}_j,\forall j\in N_{\mathcal{T}}\setminus L_{\mathcal{T}}$;
\FOR{each batch}
\STATE Assign each training sample into the leaf nodes of $\mathcal{T}$ by matching $(T_i,D_i)$ to the ordinal ranks of $l_k,\forall k$;
\STATE Assign $(X_i,D_i,T_i)$ to classifiers along corresponding path $l_k$;
\STATE Compute the loglikelihood of $(X_i,T_i)$ belonging to path $l_k$ by adding $D_i$ to Eqn. \ref{eqn:debiased_loglikelihood}
\STATE Compute $E(T|X,D,\mathcal{T})$ and $Var(T|X,D,\mathcal{T})$ as Eqn. \ref{eqn:debiased_expected_watchtime_tree} and Eqn. \ref{eqn:debiased_var_watchtime_tree};
\STATE Compute the final objective function $\mathcal{L}(T,X,\mathcal{T})$ as Eqn. \ref{eqn:debiased_objective_function};
\STATE Update $\mathcal{M}_j,\forall j$ by minimizing $\mathcal{L}(T,X,D,\mathcal{T})$;
\ENDFOR
\end{algorithmic}
\end{algorithm}
The training process is illustrated in Alg. \ref{alg:TPM_tree_debiased}.


\subsection{Model Architecture}
\begin{figure}
    \centering
    \includegraphics[width=0.42\textwidth]{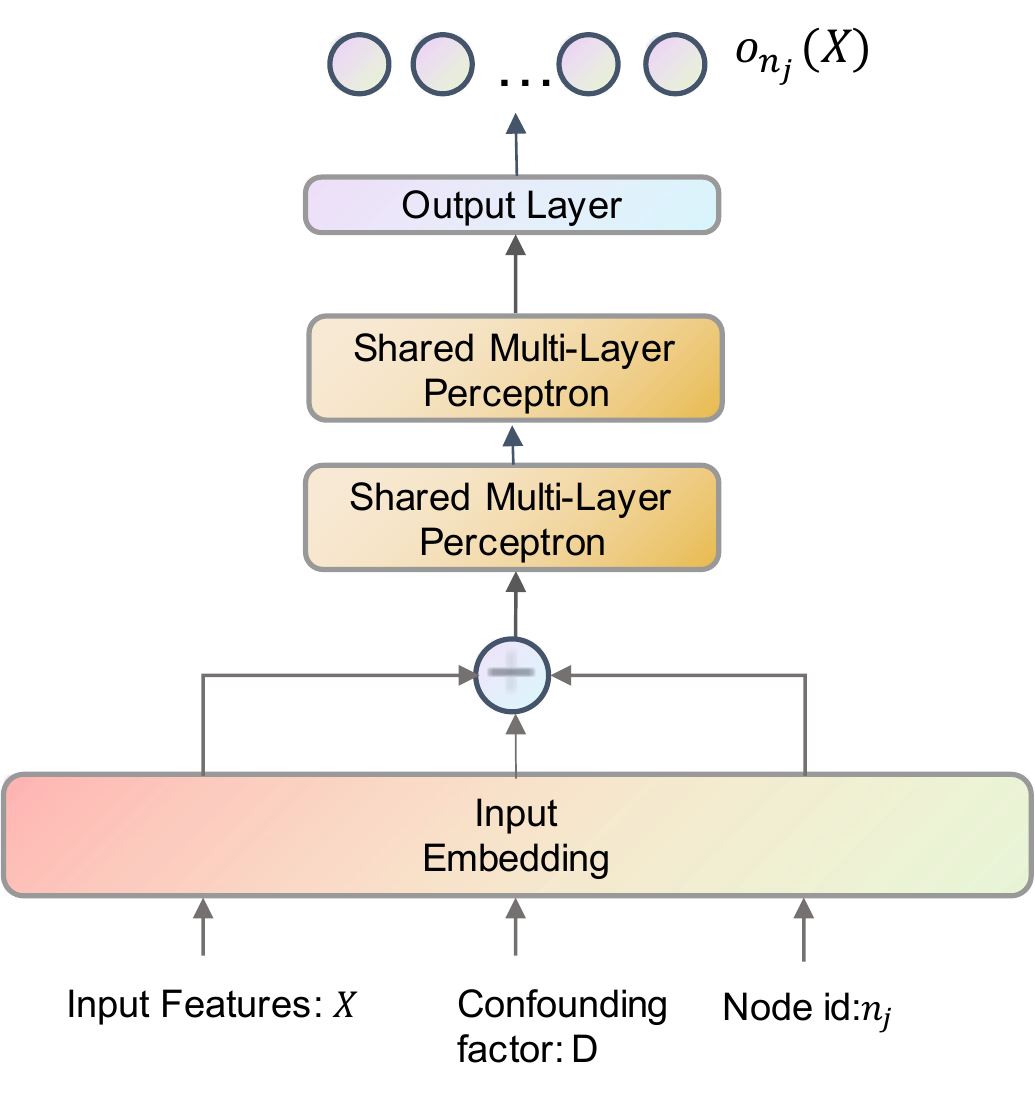}
    \caption{Network Architectures of the Classifier in TPM, where $o_{n_j}$ is the output for the task assigned to node $n_j$}
    \label{fig:network_architecture}
\end{figure}

Notice that TPM does not limit the architecture of classifiers, any architecture for binary classifier applies to TPM. Therefore we adopt a multiple layer perceptron as the backbone structure for the classifiers. The architecture is presented in Fig \ref{fig:network_architecture}.


Since each non-leaf node in a tree corresponds to a binary classification task, a naive design is to build one classifier for each node where the classifiers are trained independently. However this would cause a considerable large model size thus does not apply to the real-life environment. Therefore we design a single model for all classification tasks by sharing parameters of hidden layers across tasks. Meanwhile, task-specific output layers are introduced into the network to produce outputs for each node.

\section{Experiments} 
\label{experiments}
We conduct extensive experiments in both offline and online environments to demonstrate the effectiveness of TPM. Three research questions are investigated in the experiments:
\begin{itemize}
    \item First, how does TPM perform in comparison with state-of-the-art methods for watch time prediction in terms of recommendation accuracy?
    \item Second, how does TPM perform when combined with backdoor adjustment?
    \item Third, how do tree construction in TPM and variance modeling affect its performance? 
\end{itemize}

\subsection{Experiment Setup}
Now we provide an introduction to the experiment setup, including dataset, methods for comparison and metrics for evaluation.
\subsubsection{Datasets}
We adopt two public datasets Kuaishou (collected from Kuaishou App \footnote{https://kuairec.com/}) and CIKM16 (from CIKM16 Cup) for offline experiments. Note that CIKM16 aims to predict the dwell time for each session in online search results. We use each item in the session as a single feature for input.
Kuaishou dataset contains $7,176$ users, $10,728$ items, $12,530,806$ impressions; and CIKM16 dataset contains $310,302$ sessions, $122,991$ items and the average length of each session is $3.981$.


\subsubsection{Methods}

Two state-of-the-art methods for watch time prediction are selected for comparison, including WLR (Weighted Logistic Regression) and D2Q (Duration-Deconfounded Quantile). Moreover, ordinal regression is a method for transforming regression to classification, and it is also selected for comparison. The details of these methods are presented as follows:

\begin{itemize}
    \item WLR (Weighted Logistic Regression) \cite{covington2016deep}: This method treats watch time regression as binary classification problem, where impressed and clicked videos are positive samples and impressed but unclicked ones are negative samples. The losses of positive samples are weighted with watch time and the learned odds are used as approximate watch time. However there is no explicit negative samples in full-screen video streaming apps since all videos are impressed and played. We follow the implementation in D2Q \cite{zhan2022deconfounding} by treating short-played samples as negative ones.
    \item D2Q (Duration-Deconfounded Quantile): This approach first splits samples into ten groups based on the duration of videos. Then a regression model is trained to predict the watch time quantile for each group. Finally a prediction of watch time is retrieved by mapping predicted quantile to the watch time domain. In our experiments, the duration group number is set to 32, which achieves the best performances in predictive accuracy.
    \item OR (Ordinal Regression) \cite{NIPS2001_5531a583}: Ordinal regression transforms labels into K ranks and each rank is assigned with a classifier predicting whether the prediction is greater than the rank. As no existing studies on watch time prediction have ever adopted ordinal regression for modeling, we build a baseline for comparison by applying ordinal regression to watch time prediction directly. Meanwhile, we introduce deconfounding factors into this method in a same way with D2Q and TPM, for fair comparison.
    \item TPM\footnote{https://github.com/jackielinxiao/TPM}: The proposed approach in this paper. Since D2Q focuses on duration bias in recommendation, we also set confounding factor to video duration for comparison.
\end{itemize}
For fair comparison, the model structures of these approaches are the same except the output layers and corresponding loss functions.

\subsubsection{Metrics}
As we concern with an accurate prediction of watch time as well as its ranking capability, we adopt two metrics for evaluation, including MAE (Mean Absolute Error) and XAUC \cite{zhan2022deconfounding}:

\begin{itemize}
    \item MAE (Mean Average Error): This metric is a typical measurement for evaluating regression accuracy. Denote the predition as $\hat{y}$ and the true watch time as $y$, 
    \begin{displaymath}
        MAE = \frac{1}{N}\sum_{i=1}^N|\hat{y}_i-y|
    \end{displaymath}
    \item XAUC\cite{zhan2022deconfounding}: this metric evaluates if the predictions of two samples are in the same order with their true watch time. Such pairs are uniformly sampled and the percentile of samples that are correctly ordered by predictions is XAUC.
\end{itemize}

\subsection{Offline Experiments}

\subsubsection{Comparison with other methods}
\begin{table}
  \caption{Comparison between TPM and other approaches}
  \label{tab:offline_exp_comparison}
  \begin{tabular}{ccccc}
    \toprule
    \multirow{2}{*}{Approaches} & \multicolumn{2}{c}{Kuaishou} & \multicolumn{2}{c}{CIKM16}\\
    
    \multirow{2}{*}{} & MAE & XAUC & MAE & XAUC\\
    \midrule
    WLR & 6.047 & 0.525 & 0.998 & 0.672 \\
    D2Q & 5.426 & 0.565 & 0.899 & 0.661 \\
    OR & 5.321 & 0.558 & 0.918 & 0.664 \\
    TPM & $\mathbf{4.741}$ & $\mathbf{0.599}$ & $\mathbf{0.884}$ & $\mathbf{0.676}$ \\
  \bottomrule
\end{tabular}
\end{table}

We compare the performances of different approaches and the results are listed in Table \ref{tab:offline_exp_comparison}. Notice that TPM achieves the superior performances over other approaches significantly, this verifies the advantage of TPM in predictive accuracy. The comparison between TPM, OR and D2Q indicates the effectiveness of introducing ordinal relationships into watch time prediction. Meanwhile the comparison between TPM and OR further verifies the benefits of modeling conditional dependence in watch time prediction.

\subsubsection{Ablation Studies}

\begin{table}
  \caption{TPM with different components on KuaiRec}
  \label{tab:offline_exp_comparison_ablation}
  \begin{tabular}{ccccc}
    \toprule
    \multirow{2}{*}{Approaches} & \multicolumn{2}{c}{Watch Time} & \multicolumn{2}{c}{Watch Time Ratio}\\
    
    \multirow{2}{*}{} & MAE & XAUC & MAE & XAUC\\
    \midrule
    TPM & 4.741 & 0.599 & 0.467 & 0.712 \\
    TPM w.t.o. mse & 4.887 & 0.595 & 0.482 & 0.711 \\
    TPM w.t.o. var & 4.875 & 0.592 & 0.480 & 0.706 \\
    TPM w.t.o. deconfounding & ${4.984}$ & ${0.576}$ & ${0.497}$ & ${0.691}$ \\
    OR w.t.o. deconfounding & ${5.312}$ & ${0.549}$ & ${0.517}$ & ${0.673}$ \\

  \bottomrule
\end{tabular}
\end{table}
We conduct ablation studies on TPM and the results are listed in Table \ref{tab:offline_exp_comparison_ablation}.
The comparison between TPMs with/without mse loss indicate that adding mse loss helps to improve the metric of MAE without sacrificing too much on the ranking metric. The comparison between TPMs with/without var loss indicate that adding variance constraints helps to improve the accuracy. And TPMs with/without deconfounding factors indicate that video watching behaviors are indeed easily affected by the factors, and TPM without deconfounding factors is still competitive because of its considering of the ordinal relationships, the conditional dependency and the variances.

\pgfplotsset{
axis background/.style={fill=gallery},
grid=both,
  xtick pos=left,
  ytick pos=left,
  tick style={
    major grid style={style=white,line width=1pt},
    minor grid style=bgc,
    draw=none
    },
  minor tick num=1,
  ymajorgrids,
  major grid style={draw=white},
  y axis line style={opacity=0},
  tickwidth=0pt,
}


\begin{figure}[!htbp]
\centering
    \begin{tikzpicture}[scale=0.48]
  \begin{groupplot}[
      group style={group size=2 by 1,
          horizontal sep = 40pt}, 
        xticklabels={4, 8, 16, 32},
        xtick={1, 2, 3, 4},
        ymajorgrids,
        major grid style={draw=white},
        y axis line style={opacity=0},
        tickwidth=0pt,
        yticklabel style={
        /pgf/number format/fixed,
        /pgf/number format/precision=5
        },
        scaled y ticks=false,
        every axis title/.append style={at={(0.1,0.8)},font=\bfseries}
      ]
    \nextgroupplot[
   xlabel=$\#$Duration Group,
        ylabel=XAUC,
        legend pos=south east,
        ymin = 0.50, ymax = 0.60
    ]
    \addplot[thick,color=black,mark= otimes] coordinates {
          (1,0.567)
          (2,0.575)
          (3,0.575)
          (4,0.585)
          
        };\addlegendentry{TPM}
          
        \addplot[thick,color=violet,mark=*] coordinates {
          (1,0.525)
          (2,0.525)
          (3,0.525)
          (4,0.525)
          
        };\addlegendentry{WLR}

        \nextgroupplot[xlabel=$\#$ Duration Group,
        ylabel=MAE,
        legend pos=south east,
        ymin = 4.0, ymax = 6.2
        ]
        \addplot[thick,color=black,mark=otimes] coordinates {
          (1,4.981)
          (2,4.909)
          (3,4.889)
          (4,4.812)
        };\addlegendentry{TPM}
        \addplot[thick,color=violet,mark=*] coordinates {
          (1,6.047)
          (2,6.047)
          (3,6.047)
          (4,6.047)
        };\addlegendentry{WLR}

  \end{groupplot}
\end{tikzpicture}

    \caption{The performances of TPM with various numbers of groups when the numer of leafs is 32}
    \label{fig:dur_groups}
\end{figure}
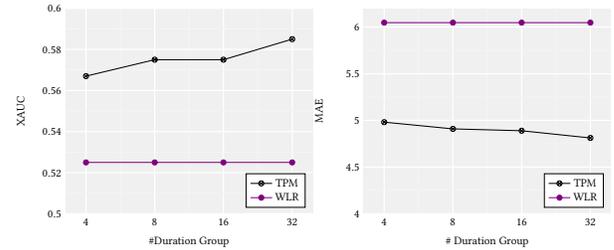

  
\pgfplotsset{
axis background/.style={fill=gallery},
grid=both,
  xtick pos=left,
  ytick pos=left,
  tick style={
    major grid style={style=white,line width=1pt},
    minor grid style=bgc,
    draw=none
    },
  minor tick num=1,
  ymajorgrids,
  major grid style={draw=white},
  y axis line style={opacity=0},
  tickwidth=0pt,
}


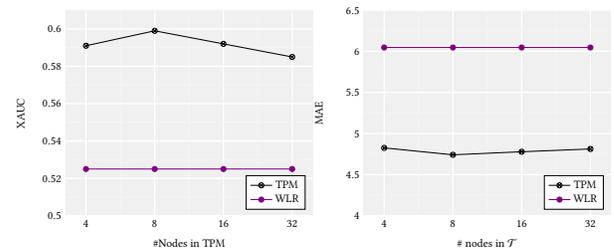
\begin{figure}[!htbp]
\centering
    \begin{tikzpicture}[scale=0.48]
  \begin{groupplot}[
      group style={group size=2 by 1,
          horizontal sep = 40pt}, 
        xticklabels={4, 8, 16, 32},
        xtick={1, 2, 3, 4},
        ymajorgrids,
        major grid style={draw=white},
        y axis line style={opacity=0},
        tickwidth=0pt,
        yticklabel style={
        /pgf/number format/fixed,
        /pgf/number format/precision=5
        },
        scaled y ticks=false,
        every axis title/.append style={at={(0.1,0.8)},font=\bfseries}
      ]
    \nextgroupplot[
   xlabel=$\#$Nodes in TPM,
        ylabel=XAUC,
        legend pos=south east,
        ymin = 0.50, ymax = 0.61
    ]
    \addplot[thick,color=black,mark= otimes] coordinates {
          (1,0.591)
          (2,0.599)
          (3,0.592)
          (4,0.585)
          
        };\addlegendentry{TPM}
          
        \addplot[thick,color=violet,mark=*] coordinates {
          (1,0.525)
          (2,0.525)
          (3,0.525)
          (4,0.525)
          
        };\addlegendentry{WLR}

        \nextgroupplot[xlabel= $\#$ nodes in $\mathcal{T}$,
        ylabel=MAE,
        legend pos=south east,
        ymin = 4.0, ymax = 6.5
        ]
        \addplot[thick,color=black,mark=otimes] coordinates {
          (1,4.825)
          (2,4.741)
          (3,4.778)
          (4,4.812)
        };\addlegendentry{TPM}
        \addplot[thick,color=violet,mark=*] coordinates {
          (1,6.047)
          (2,6.047)
          (3,6.047)
          (4,6.047)
        };\addlegendentry{WLR}

  \end{groupplot}
\end{tikzpicture}

    \caption{The performances of TPM with various numbers of nodes in the decomposition tree.}
    \label{fig:nodes_tpm}
\end{figure}

  
We alter the number of duration groups and the performances of TPM are depicted in Fig. \ref{fig:dur_groups}. And the results indicate that splitting samples by duration indeed helps and TPM can seamlessly accommodate with backdoor adjustments.
We also conduct experiments of TPM with various number of nodes in the tree, the results are illustrated in Fig.\ref{fig:nodes_tpm}. As depicted in the figure, there is a proper number of nodes for tree construction. This coincides with the intuition that the tree should be constructed according to the task and dataset.
Meanwhile, to illustrate the effects of variance for uncertainty modeling, we alter the weights for variance in the loss function and present the results in Fig. \ref{fig:variances_tpm}. The results reveal that there is a proper weight of variance which leads to lower uncertainty and satisfactory accuracy.

\pgfplotsset{
axis background/.style={fill=gallery},
grid=both,
  xtick pos=left,
  ytick pos=left,
  tick style={
    major grid style={style=white,line width=1pt},
    minor grid style=bgc,
    draw=none
    },
  minor tick num=1,
  ymajorgrids,
  major grid style={draw=white},
  y axis line style={opacity=0},
  tickwidth=0pt,
}


\begin{figure}[!htbp]
\centering
    \begin{tikzpicture}[scale=0.48]
  \begin{groupplot}[
      group style={group size=2 by 1,
          horizontal sep = 40pt}, 
        xticklabels={0.1, 0.5, 1.0, 10.0},
        xtick={1, 2, 3, 4},
        ymajorgrids,
        major grid style={draw=white},
        y axis line style={opacity=0},
        tickwidth=0pt,
        yticklabel style={
        /pgf/number format/fixed,
        /pgf/number format/precision=5
        },
        scaled y ticks=false,
        every axis title/.append style={at={(0.1,0.8)},font=\bfseries}
      ]
    \nextgroupplot[
   xlabel= Weight of uncertainty loss,
        ylabel=XAUC,
        legend pos=south east,
        ymin = 0.50, ymax = 0.61
    ]
    \addplot[thick,color=black,mark= otimes] coordinates {
          (1,0.580)
          (2,0.582)
          (3,0.588)
          (4,0.581)
          
        };\addlegendentry{TPM}
          

        \nextgroupplot[xlabel= Weight of uncertainty loss,
        ylabel=Sqrt of Variance,
        legend pos=south east,
        ymin = 0.0, ymax = 0.1
        ]
        \addplot[thick,color=black,mark=otimes] coordinates {
          (1,0.0667)
          (2,0.0595)
          (3,0.0523)
          (4,0.0485)
        };\addlegendentry{TPM}

  \end{groupplot}
\end{tikzpicture}

    \caption{The performances of TPM with different uncertainty weight in training when the num of leafs is 32.}
    \label{fig:variances_tpm}
\end{figure}
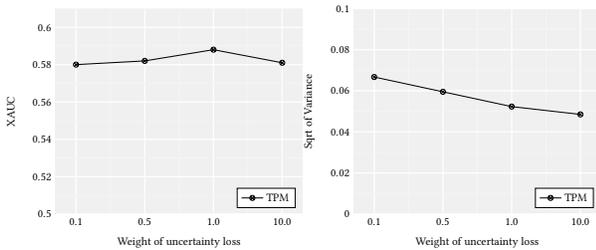

  
\subsection{Online Experiments}
We also conduct online A/B experiments on a real-world short-video recommender system in KuaiShou APP. As D2Q is a state-of-the-art method for watch time prediction, it is adopted as a baseline for comparison. For TPM, the decomposition tree is design as a complete binary tree with 32 leaf nodes, and the number of duration groups is set to 32.
\subsubsection{Experiment Setup}

In online A/B experiments, the traffic is split into ten buckets uniformly. Two buckets of traffic are assigned to baseline while the other two are assigned to TPM. As revealed in \cite{zhan2022deconfounding}, Kuaishou serves over 320 million users daily and the results collected from $20\%$ of traffic is very convincing.

The real-life recommender systems are usually complicated. However, most of the systems follow a two-stage framework where a set of candidate items are retrieved in the first stage and the top-ranking items are selected from the candidates in the ranking stage. Watch time prediction serves as one component in the ranking stage. The items are ranked with multiple predictions (including watch time predictions) and those with higher watch time predictions are more likely to be recommended.

\subsubsection{Experiment Results}
The experiments have been launched on the system for 4 days, and the results are listed in Table \ref{tab:online_exp_comparison}. 
\begin{table}
  \caption{Comparison between TPM and baseline online, all values are the relative improvements of TPM over the baseline.
  Watch Time and Forward are positive metrics where higher values are better; Short View is a negative metric where lower values are better. Meanwhile Forward is a constraint metric, and an experiment with more than $1\%$ drop of constraint metrics is not acceptable. For Online A/B tests, an improvement of $0.1\%$ in watch time is very significant.} 
  \label{tab:online_exp_comparison}
  \begin{tabular}{cccc}
    \toprule
    Days &  Watch Time & Forward & Short View\\
    \midrule
    Day 1 & $\mathbf{0.246\%}$ & $-0.002\%$ & $\mathbf{-0.312\%}$ \\
    Day 2 & $\mathbf{0.210\%}$ & $0.040\%$ & $\mathbf{-0.139\%}$ \\
    Day 3 & $\mathbf{0.234\%}$ & $0.001\%$ & $\mathbf{-0.110\%}$ \\
    Day 4 & $\mathbf{0.265\%}$ & $0.225\%$ & $\mathbf{-0.220\%}$ \\
  \bottomrule
\end{tabular}
\end{table}
The metrics for online experiments include accumulated watch time, forward counts (forward the video to friends) and short view counts (watch time is short in respect to the video duration). In online experiments, watch time is a core metric while forward is a constrained metric.
Notice that TPM outperforms the baseline in watch time related metrics which verifies the advantage of TPM in predictive accuracy. Moreover the number of negative feed-backs are significantly lower in TPM. And this coincides with the idea of modeling uncertainty in TPM, aiming to produce both accurate and confident predictions.
Meanwhile, the differences between TPM and baseline on metrics of interaction are insignificant, thus can be neglected safely.


\section{Conclusion}
Watch time prediction is one of the core problems in short-video recommendation, as its accuracy affects the quality of videos recommended to users, thus impacting user engagement to the platform. We point out that four issues should be addressed in a real-world watch time prediction framework: first, the ordinal differences between watch time values should be considered; second, the conditional dependence between the video-watching behaviors should be modeled; third, the uncertainty of predictions should be involved in the framework; forth, the framework should take bias amplification into consideration.

To solve these issues simultaneously, we propose TPM (Tree-based Progressive regression Model) for watch time prediction. We reveal that watch time prediction can be decomposed into several conditional dependent classification problems that are organized into a tree structure. Meanwhile, the variance of watch time predictions is introduced into the objective function as model uncertainty. And the bias amplification problem is addressed by incorporating backdoor adjustment into TPM seamlessly. 

Extensive offline evaluations and online experiments in real-life recommender systems have been conducted and the results validate the effectiveness of TPM. Moreover, TPM has already been deployed in Kuaishou APP, serving over $300$ million users daily.

\bibliographystyle{ACM-Reference-Format}
\bibliography{sample-sigconf}
\end{document}